# Gravitational Annealing of Colloidal Crystals

Yoshihisa Suzuki[1,*], Jin Endoh[2], Atsushi Mori[3], Tomoki Yabutani[1] and Katsuhiro Tamura[1]

[1]Department of Life System, Institute of Technology and Science, The University of Tokushima, Tokushima, Japan

[2]Department of Chemical Science and Technology, The University of Tokushima, Tokushima, Japan

[3]Department of Advanced Materials, Institute of Technology and Science, The University of Tokushima, Tokushima, Japan

[*]suzuki@chem.tokushima-u.ac.jp

**Abstract.** A silica colloidal crystal obtained by centrifugation at 9 G for 2 days in water was annealed by additional stronger centrifugation at 50 G for 5 days. The number of the striations observed in the colloidal crystal under crossed polarized light decreased at some parts in a growth container after the additional centrifugation, while the number also increased at the other parts. The decrease probably shows the shrinkage of the stacking disorders under high gravity, while the increase probably shows the production of new stacking disorders.

**Introduction**

A close-packed colloidal crystal is useful as a template for inverse opals with perfect three-dimensional photonic bandgap [1], and it is easily fabricated by drying highly-concentrated colloidal crystals [2,3,4]. We successfully fabricated large grains of the highly-concentrated colloidal crystals by centrifugation. However, we have not controlled lattice defects in the grains yet.

Although defect control in colloidal crystals has been one of the most important subjects, and many studies have been done, they still have some problems. Most of the efforts have been dedicated to control crystallization processes using square-shaped templates not to induce stacking faults into face-centered cubic (FCC) colloidal crystals [5,6]. Although these efforts have achieved a single colloidal crystal with square-centimeter area and hundreds layers height, we cannot apply further improvements on the obtained crystal. On the other hand, only a few studies show annealing of the obtained crystals [7]. Harada, et al. showed that an optical tweezer could melt a colloidal crystal locally, and they could refine the particle arrangements regularly in the crystals. However, their method can be applied to a limited small area.

We tried to apply gravitational force to anneal the obtained colloidal crystals by centrifugation. Centrifugation would be the best choice for annealing the crystals, since we have already fabricated large grains by centrifugation, and gravitational forces is known to reduce stacking disorders in hard-sphere crystals by Monte Carlo simulations [8,9]. In addition, centrifugal forces are applied to all the parts of the crystal macroscopically.

In this study, we tried to reduce the number of stacking disorders in colloidal crystals by centrifugation.

**Experimental**

Aqueous suspensions of silica particles (Nippon Shokubai Seahoster KE-W10, $d$ = 110 nm, volume fraction of the dispersion $\varphi$ = 0.0968) were used without further purification. The suspensions were transferred into a hand-made observation container. The container has an inverted-triangle shape (Fig. 1). The container has an inverted-triangle internal-shaped spacer (vertical-angle = 60°, maximum width = 6 mm, thickness = 0.2 mm) between two glass slides, and the container was set on a sample holder of a stirrer (Iuchi, SM 101). Centrifugation of the particles was conducted by rotating the holder. Obtained crystals were characterized using a transmitted-light polarization microscope (Olympus, VANOX). We observed retardation colors of the grains [5].

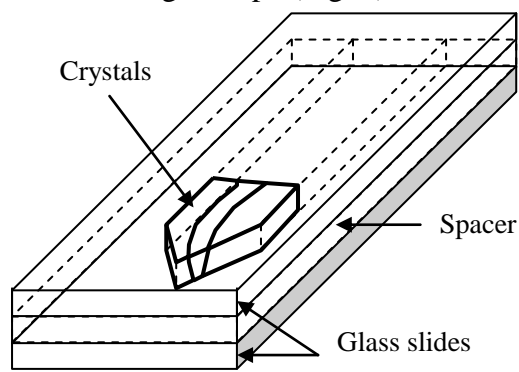

Fig. 1. Schematic illustration of an observation container with inverted-triangle internal-shaped spacer.

**Results and Discussion**

**Colloidal crystallization by centrifugation.** A highly concentrated colloidal crystal was obtained by centrifugation at 9 G for 2 days. An obtained crystal is shown in Fig. 2. As previously reported [2,3,4], grains in the crystal have columnar shapes from the bottom of a growth container to the growth front. A polarized transmission image of a grain in the crystal showed some striations in the grain. The striations probably show the stacking disorders in the grain [5], since striations in the same grain seem to be arranged almost parallel each other. Stacking disorders in a crystal are generally produced by growth faults and the increase of stress. Thus, we tried to anneal the disorders by applying higher centrifugal forces to obtained crystals.

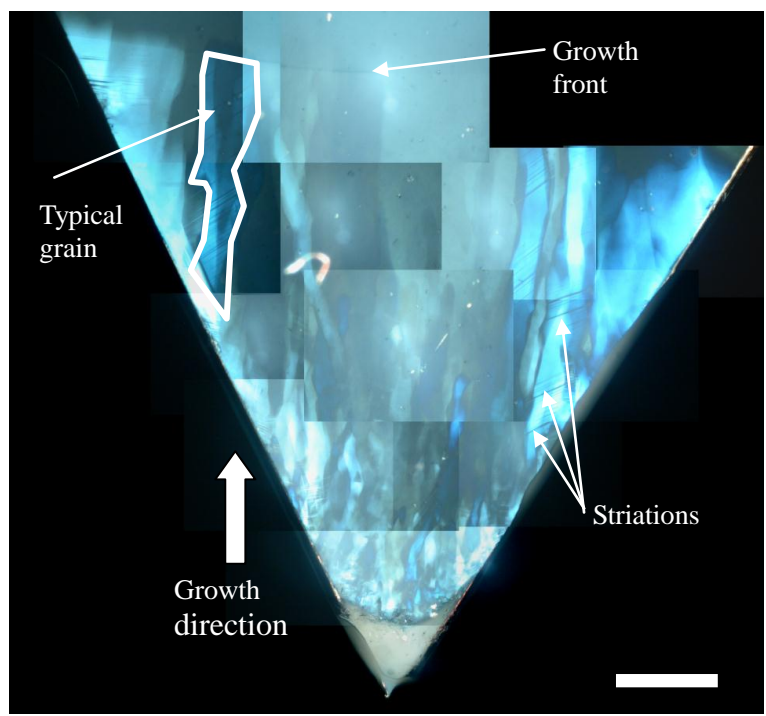

Fig. 2. Merged polarized transmission images of a silica colloidal crystal which was obtained by centrifugation at 9 G for 2 days in an inverted-triangle internal-shaped container. Different retardation colors and contrasts in different grains show the difference in their crystallographic direction. Some grains contain striations inside. Scale bar indicates 1 mm.

**Change in the size and shape of grains after a gravitational annealing process.** After the first observation, the crystal was centrifuged again at 50 G for 5 days (Fig. 3). During the first microscopic observation, the interface between the crystal and dispersion dissolved a little. After that, we applied an additional centrifugal acceleration, and obtained a crystal as shown in Fig.3 (b).

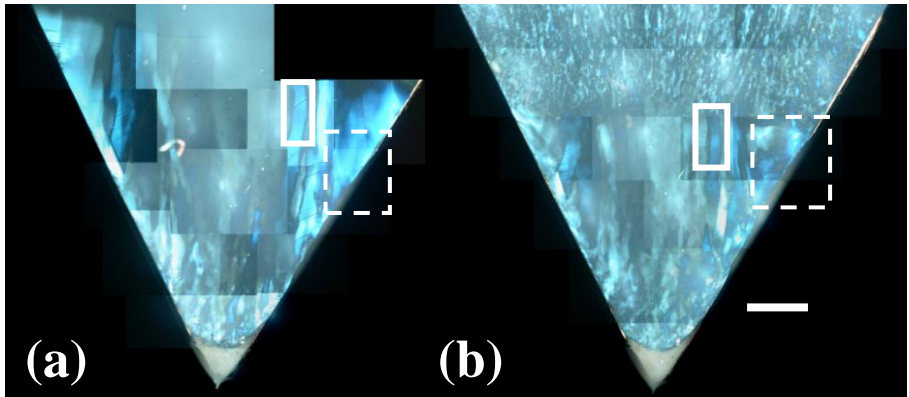

Fig. 3. A silica colloidal crystal (a) before and (b) after an additional centrifugation for gravitational annealing (at 50 G for 5 days). Scale bar indicates 1 mm.

In Fig. 3 (b), there is a border between the upper part with smaller grains and the lower part which is already shown in Fig. 3 (a). Smaller grains are nucleated during the additional centrifugation (at 50 G for 5 days). Thus, the border indicates the dissolution interface of a firstly obtained crystal (Fig. 3 (a)). In the lower part in (b), all grains seem to get smaller than those shown in Fig. 3 (a). In addition, the shape of the grains changes a little after the additional centrifugation process. This is due to the gravitational force applied during the annealing process. Thus, we name this process a gravitational annealing process.

**Shrinkage of stacking disorders.** The number of striations in a grain decreased after the above gravitational annealing process (Fig. 4). This is probably due to gliding of triangular lattice planes in a face-centered cubic (FCC) colloidal crystal. Mori et al. [8] reported that Monte Carlo simulation results of the crystallization of hard-sphere colloidal crystals show the shrinkage of stacking faults with increasing gravitational acceleration. Mori, et al. also showed that the cross coupling term between the elastic field due to gravity and that due to a Shockley partial dislocation in a hard-sphere colloidal crystal is an increasing function of the depth of the Shockley partial dislocation [9].

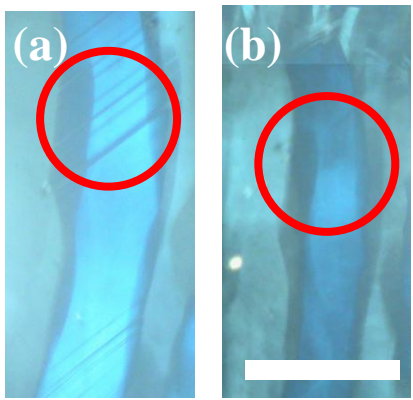

Fig. 4. Disappearance of striations in a grain. Both (a) and (b) areas are also shown in Fig. 3 (a) and (b), respectively, as squares with white solid lines. After a gravitational annealing, the striations in the circle of (a) disappeared in that of (b). Scale bar indicates 500 μm.

**Production of new stacking disorders.** The number of striations in another grain increased after the same gravitational annealing process (Fig. 5). As a result, both shrinkage and production of striations occur during the same gravitational annealing process. Such different results are mainly due to different direction of striations. We should collect much more data, and clarify the correlation between the directions and the number of striations in future.

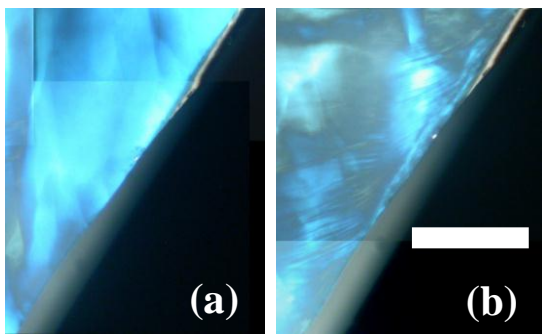

Fig. 5. Production of new striations in a grain. Both (a) and (b) areas are also shown in Fig. 3 (a) and (b), respectively, as squares with white dashed lines. Black areas indicate those of a spacer. After a gravitational annealing, new striations near the side wall of the spacer are produced in (b). Scale bar indicates 500 μm.

**Conclusions**

In this study, we observed inner structures of a silica colloidal crystal using a polarized transmittance microscope before and after a gravitational annealing. Key results found in this paper are as follows.
 (1) A silica colloidal crystal was obtained by centrifugation at 9 G for 2 days in an inverted-triangle internal-shaped container. Some grains contain striations inside. The striations indicate stacking disorders.
 (2) The colloidal crystal shrunk after the additional centrifugation at 50 G for 5 days (gravitational annealing). The shape of each grain deformed a little during the gravitational annealing.
 (3) Striations in a grain disappeared during the gravitational annealing. The disappearance indicates annealing-out of stacking disorders. This is probably due to gliding of triangle planes in a face-centered cubic colloidal crystal by applying gravity.
 (4) However, in the other grains, striations are newly produced. Disappearance and production probably depend on the relations between the direction of stacking disorders and gravity.